\documentclass[aps,pre,twocolumn,showpacs,amssymb]{revtex4}

\usepackage[pdftex]{graphicx}
\usepackage{dcolumn}
\usepackage{bm}
\def\be{\begin{equation}}
\def\en{\end{equation}}

\newcommand{\av}[1]{\langle{#1}\rangle}
\newcommand{\AV}[1]{{\bigg\langle}{#1}{\bigg\rangle}}
\def\gs{\gtrsim}
\def\ls{\lesssim}
\newcommand{\bi}[1]{\mbox{\boldmath$#1$}}

\def\bea{\begin{eqnarray}}
\def\ena{\end{eqnarray}}
\def\a{_{\alpha\beta}}

\begin{document}


\title{
Slow relaxations and stringlike jump  motions  
in fragile   glass-forming liquids: 
Breakdown of the Stokes-Einstein relation  
  }



\affiliation{Department of Physics, Kyoto University, Kyoto 606-8502,
Japan}


\author{Takeshi  Kawasaki and Akira Onuki}

\affiliation{Department of Physics, Kyoto University, Kyoto 606-8502,
Japan}


\date{\today}

\begin{abstract}
We perform molecular dynamics simulation on  a glass-forming liquid 
binary mixture  with  the soft-core potential  in three dimensions. 
We  investigate 
 crossover of the 
configuration changes caused by stringlike jump  motions.  
With lowering   the temperature $T$,  
  the motions of   the particles 
composing   strings become larger 
in sizes  and displacements, while  
 those of the particles  surrounding  strings 
become smaller. Then, 
the  contribution of the latter to time-correlation functions 
tends to be  long-lived  as $T$ is lowered. 
As a result,  the relaxation time $\tau_\alpha$   
 and the viscosity $\eta$ 
grow more steeply  than the  inverse diffusion constant $D^{-1}$ 
at low $T$,  leading   to  breakdown 
of the Stokes-Einstein  relation. 
At low $T$,  the  diffusion  occurs 
as activation processes and 
may well be described by short-time 
 analysis of rare  jump motions with 
broken bonds and large displacements. 
Some  characteristic  features  of the van Hove 
self-correlation  function  arise from escape 
jumps over high potential barriers. 
\end{abstract}



\maketitle


\section{Introduction}
\setcounter{equation}{0}

Recently, much attention has been paid to   the slow dynamics 
in glass-forming liquids \cite{Binder}, where 
the relaxation time $\tau_\alpha$ 
becomes  extremely long  and the 
viscosity $\eta$ grow dramatically 
with lowering the temperature $T$. 
In particular, it  is well known experimentally 
 that   the curve of $\log \eta$ vs $1/T$ 
 (the Angell plot\cite{Angel}) becomes  steeper 
with increasing  $1/T$  deviating from the  Arrhenius
behavior for fragile glass-forming liquids, while 
the Arrhenius behavior has been  observed for strong glass 
formers like silica. 
However, it is not yet well understood how this dynamical 
crossover takes place  microscopically for fragile glass 
formers.  Hence, in this paper, we aim   to investigate this aspect 
using molecular dynamics simulation.

In fragile glass-formers,  stringlike  motions  of mobile 
particles \cite{Sc,Kob,Glo} play a major role in 
the structural relaxation. The distribution of 
 string lengths \cite{Kob} was shown to widen with lowering $T$. However,  
 the  behaviors and the roles of stringlike motions  
have not yet been fully  disclosed  in the structural relaxation 
and the plastic deformations. In this paper, we analyze   the configuration 
changes   with  the bond breakage theory \cite{yo1,yo}, which was 
  originally used to detect the dynamic heterogeneity  \cite{Harrowell,book}
and has recently been generalized as a statistical-mechanical theory 
of irreversible particle rearrangements  \cite{SKO}.  
We  then confirm   that the configuration   changes  mostly 
occur as stringlike  motions  of mobile 
particles at low $T$.   We shall furthermore see  that 
 the particles composing strings and those 
 surrounding them  behave  differently depending on $T$  
in a binary mixture at a fixed high density $n$ in three dimensions (3D).  
    With lowering $T$, 
the former move  over longer  distances with multiple 
broken bonds, while the latter over shorter  distances 
with a single broken bond.  
The number of the latter is several times 
 larger than that of the former  because of 
the large coordination numbers  in 3D $(\sim 10$).  
Hence, the crossover  of the 
latter motions with lowering $T$  should greatly influence   
 the relaxation behavior of 
time-correlation functions and the 
 relaxation time $\tau_\alpha$ at low $T$.

In liquid, the Stokes-Einstein relation 
$D\eta a/T=$const. between the  
diffusion constant $D$ and the viscosity $\eta$  
 has been  successfully 
applied  even for a microscopic test  particle diameter $a$. 
However, this relation is systematically violated 
in fragile supercooled liquids.  
\cite{Ediger,Silescu}.  
 Sillescu {\it et al.}\cite{Silescu} 
observed the power law behavior 
$D \propto \eta^{-\nu}$ with $\nu \cong 0.75$ at low $T$. 
The origin of this violation 
has  often  been ascribed to the dynamic heterogeneity 
 without   clear-cut calculations.  
A number of simulations have not clearly 
explained  this violation 
\cite{Harrowell,Wa,yo-diffusion,Rei,Sch,Szamel}.  
We shall see that the above-mentioned  crossover 
of the particle motions around strings 
is the main origin of the violation  
for our  fragile glass-forming  system.

 For sufficiently low $T$ in the  glass-forming condition, 
the van Hove  self-correlation function $G_s(r,t)$ 
tends to have a  minimum 
at a particle length  $r_{\rm m}$ and   slowly 
grows  in the outer region $r>r_{\rm m}$ 
forming  secondary peaks    
\cite{Sastry,Wa,yo-diffusion,Rei,Sch,Szamel}. 
Here, $r$ represents the particle displacement length 
in a time interval of $t$. 
We shall see that its  minimum value at $r=r_{\rm m}$ 
becomes very small  at low $T$, indicating fast passage of the 
particles across  potential barriers  at  $r\sim r_{\rm m}$.   
These behaviors  reflect the fact  that the 
diffusion   arises  from  
 escape  jumps from well-defined   transient cages with long life times. 
Sastry {\it et al.} 
 called this  the  landscape-dominated regime \cite{Sastry}, which  
emerges  on approaching the glass transition.  
We shall also see that 
 the contribution to the mean square displacement 
 $M(t)$ from the particles with large displacements 
behaves as $6Dt$ soon after the ballistic 
time region, where 
 the thermal vibrational contributions  from the 
other particles  in the 
interior $r<r_{\rm m}$  are removed.

The organization of this paper is as follows.
In Sec.II, our simulation method will be   explained. 
 In Sec.III, the bond-breakage 
theory  in our previous work \cite{yo,yo1,SKO} will be  
used to examine the particle motions around  strings 
for various $T$. 
In Sec.IV,  we  will examine 
the long-distance diffusion, which will turn out to be governed by 
the activated dynamics at low $T$. 
We will also  examine  the van Hove self-correlation 
function $G_s(r,t)$ in the landscape-dominated regime.

\setcounter{equation}{0}
\section{Numerical method}

In this paper we  show 
results of molecular dynamics simulation 
of  binary mixtures composed of 
two   species, 1 and 2,  
 in 3D   at   low $T$. 
 We impose  the periodic boundary condition  
without applying shear flow. 
The composition is $c=N_2/(N_1+N_2)=0.5$ 
 and the total particle number 
is $N=N_1+N_2=10^4$. 
The two species  have different diameters 
 $\sigma_1$ and $\sigma_2$ with 
$\sigma_2/\sigma_1=1.2$.  
The  particles   interact via   
the soft-core  potential,  
\begin{equation}
v_{\alpha\beta} (r) = 
\epsilon  \left(\frac{\sigma_{\alpha\beta}}{r}\right)^{12} -C_{\alpha\beta}
\quad  ~
(r<r_{\rm cut}),  
\label{eq:LJP}
\end{equation}
where  $\alpha$ and $\beta$ represent 
the particle species   $(\alpha,\beta =1,2)$, 
 $r$ is the particle  distance, and $\epsilon$  
 is the characteristic interaction energy. 
The  interaction lengths are defined by 
\be     
\sigma{\a} = 
(\sigma_\alpha +\sigma_\beta )/2. 
\en 
The potential vanishes  for   $r>r_{\rm cut}= 3\sigma_1$. The constants   
 $C_{\alpha\beta}$  ensure the continuity of the potential at 
$r=r_{\rm cut}$.   
 The masses   of the two species satisfy 
  $m_2/m_1= (\sigma_2/\sigma_1)^3$. 
The average density is given by $ 
n=N/V=0.8\sigma_1^{-3}$ as in some  previous papers 
\cite{yo,yo1,yo-diffusion,SKO,Furukawa}, where 
$V$ is the system volume.   
The system length is  $L=V^{1/3}=23.2\sigma_1$. 
Space and time are  measured in units of 
$\sigma_1$ and  
\be 
\tau_0= \sigma_1\sqrt{m_1/\epsilon}.  
\en 
The temperature $T$ is  measured 
 in units of $\epsilon/k_B$.

We started from a liquid state 
at a high temperature, 
quenched the system to a final low temperature, and 
waited  for a long time of order  $10^5$. 
  We imposed a Nose-Hoover thermostat in these steps. 
However, after this preparation of initial states, we  removed 
 the  thermostat and  integrated   the Newton equations 
under the periodic boundary condition 
in the time range $t>0$.  Thus, the particle numbers, the total volume, 
and the total energy are conserved in our simulation ($NVE$ ensemble). 
At the lowest temperature 
 $T=0.24$, the simulation time was $5\times 10^5$ for each run 
and data were  taken in the time range $5\times 10^4<t<5\times 10^5$, 
during which  we did not 
detect  any appreciable aging effects  
in various quantities such as the average potential 
energy and $F_s(q,t)$ defined below.

We are interested in supercooled  states, where the structural 
 relaxation time $\tau_\alpha$ is very long.  
In terms of   the self part of the density 
 time-correlation function,  
 \be  
 F_s(q,t)=\frac{1}{N}\sum_j \AV{\exp[
 i{\bi q}\cdot \Delta{\bi r}_j(t_0,t_0+t)]}, 
\en 
 $\tau_\alpha$ is usually  defined  at $q=2\pi\sigma_1^{-1}$  by   
\be 
F_s(q,\tau_\alpha)=e^{-1} .
\en 
Here,   $\bi q$ is the wave vector with  $q=|{\bi q}|$  and 
$\Delta{\bi r}_j(t_0,t_1)= {\bi r}_j(t_1)- {\bi r}_j(t_0)$ 
is the  displacement vector 
of particle $j$. 
 In this paper,  $\av{\cdots}$  denotes taking the average 
 over   the initial time $t_0$   of the time interval  and over 
 several  simulation runs.  This is needed for  averaging quantities 
involving two widely separated times, for which   
accurate results do  not follow  only by 
the average over all the particles  for our system size.

It is worth noting that the stress time-correlation 
function behaves similarly  to $F_s(q,t)$, both  
considerably decreasing   in the  early  stage 
due to the thermal motions (the so-called 
$\beta$ relaxation) \cite{yo,Furukawa}.   
Since  the  time integral of the stress time-correlation 
function is equal to the shear viscosity $\eta$  
in the linear response regime, 
 it is natural to expect weak $T$ dependence of  the ratio 
$\eta/ \tau_\alpha$. For the same soft-core potential (2.1), 
 the following  relation was previously obtained  \cite{yo-diffusion}:   
\be 
\eta  \cong (2\pi)^{-1}\sigma_1^{-3} k_BT\tau_\alpha ,
\en
in the range $1< \tau_\alpha<10^4$ at $n=0.8\sigma_1^{-3}$.

\setcounter{equation}{0}
 \section{Bond  breakage }
 \subsection{Stringlike   motions and their crossover}
 
We analyze the configuration changes 
in  the bond breakage scheme \cite{yo,yo1,SKO}. They  
 occur as jump motions causing  irreversible particle rearrangements.  

First, we introduce the concept of bond breakage. 
At an initial time $t=t_0$, 
  two particles $i \in \alpha$ and $j \in \beta$ ($\alpha,\beta=1,2$) 
are treated to be bonded if 
\be 
r_{ij}(t_0 )<A_1\sigma\a. 
\en     
Hereafter,  
$r_{ij}(t)= |{\bi r}_i (t)-{\bi r}_j (t)|$ is the 
distance between these particles 
at time $t$.   At a later time $t=t_1>t_0$, 
this  bond is   treated to be  broken if 
\be 
r_{ij}(t_1)>A_2 \sigma\a.
\en   
Here,  $A_1\sigma_{\alpha\beta}$ 
are  slightly larger than the first peak distances  
of the pair correlation 
functions $g_{\alpha\beta}(r)$ 
and  $A_2$ 
is somewhat larger than $A_1$.
In this paper, we set $A_1= 1.3$ and $A_2=1.7$.  
 At high densities in 3D, the bond number around each particle 
is of order 10 because it is 
on   the order of the  coordination number. 
Let $N_b(t_0)$ be the number of the   bonds defined at time $t_0$ 
and $N_b(t_0, t_0+ t)$ be the number of the  unbroken  bonds 
at later time  $t_0+t$.  Then,   
the fraction of the unbroken  or surviving bonds after time $t$ 
is given by 
\be 
F_b(t)=\av{ N_b(t_0,t_0+t)/N_b(t_0)} .
\en   
The bond breakage time $\tau_b$ is determined by \cite{yo1,yo} 
\be 
F_b(\tau_b)=e^{-1}.
\en 
Nonlinear rheology in supercooled  states \cite{yo1,Shiba}
 is governed by the bond breakage and is 
characterized by $\tau_b$.

Furthermore, we introduce   the broken bond number for each  particle $i$ 
in terms of $A_1$ and $A_2$  
  as \cite{SKO}  
\bea 
{\cal B}_i (t_0,t_1) 
&=&  \sum_j  \theta(A_1 \sigma\a - r_{ij}(t_0))\nonumber\\ 
&&\times \theta( r_{ij}(t_1)-A_2 \sigma\a ) ,   
\ena 
where  $\theta(u)$ is the step function being zero for $u\le 0$ 
and 1 for $u>0$.  
This number is   a nonnegative integer, tending 
 to  zero as  $t_1 \to  t_0$ from $A_1<A_2$ 
and increasing to $1, 2, \cdots$  upon bond breakage.   
The particles with  
${\cal B}_i (t_0,t_1)>0$ may be 
 called  $\bi B$ particles, while 
those with  ${\cal B}_i (t_0,t_1)=0$ may be called 
 non-$\bi B$ particles.   The $\bi B$ particles are  
surrounded by different  particle 
configurations at the initial and final times 
$t=t_0$ and $t_1$.

The displacements of the $\bi B$ particles 
may not be large if they are neighbors of  those  
undergoing  stringlike motions. Thus it 
 is convenient to group  the $\bi B$ 
 particles further into those with  large displacement 
 $\Delta r_i  > \ell_m$  
and those with  small displacement 
 $\Delta r_i  < \ell_m$,  where   $\ell_m$ is 
  a minimum jump length and 
\be 
\Delta r_i= \Delta r_i(t_0, t_1) = |{\bi r}_i(t_1)- 
{\bi r}_i(t_0)| 
\en 
represents the displacement  in  time interval $[t_0,t_1]$. 
 These particles  are   called 
$\bi{BL}$  and $\bi{BS}$ particles, respectively. 
Their   broken bond numbers  
 are written as 
\bea
&&\hspace{-3mm}{\cal B}_i^> (t_0,t_1) ={\cal B}_i (t_0,t_1) 
\theta( \Delta r_{i}(t_0,t_1)-\ell_m) ,\\
&&\hspace{-3mm}{\cal B}_i^< (t_0,t_1) ={\cal B}_i (t_0,t_1) 
\theta(\ell_m- \Delta r_{i}(t_0,t_1)) . 
\ena 
In this paper,  we set $\ell_m =0.8$. A small change of its value 
does not essentially change our conclusions. See SubsecIVD and Fig.12 
 on this point.

\begin{figure}[t]
\includegraphics[width=1\linewidth]{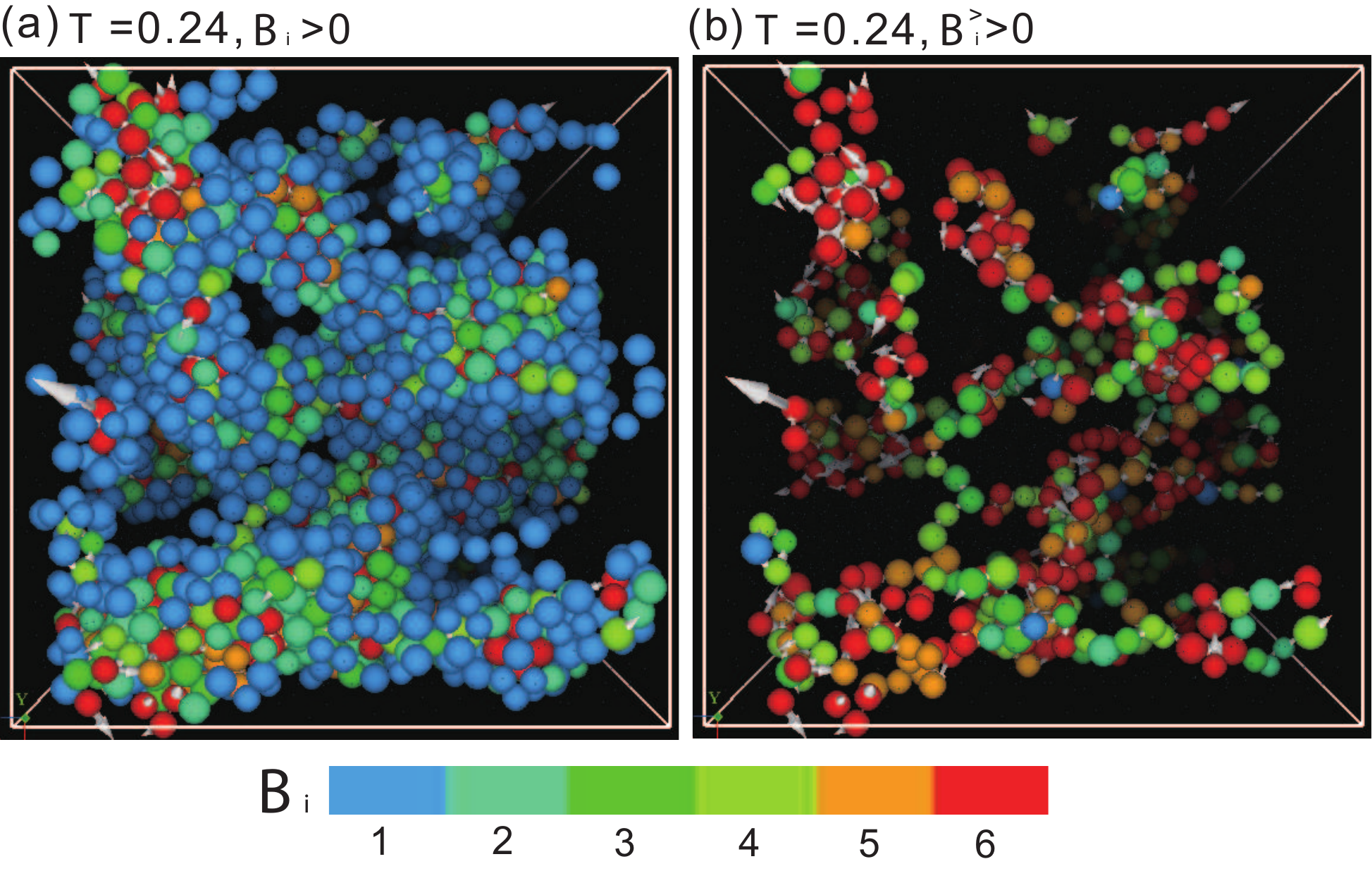}
\caption{(Color online) (a)  $\bi B$ particles (those with ${\cal B}_i= 
{\cal B}_i(t_0,t_1)>0$)  
composed of their strings and their  neighbors. 
(b) $\bi{BL}$ particles (those  with ${\cal B}_i^>(t_0,t_1)>0$)   
composed of strings, which have  large ${\cal B}_i\ge 3$ 
due to  large displacements. These are snapshots 
at  $t=t_1-t_0=10^4\cong 10^{-1}\tau_\alpha$ or at  $\phi_b^>(t)=0.1$ 
for $T=0.24$ (see Table 1).  
The particle colors represent  ${\cal B}_i$ 
according to the color bar below. 
Arrows represents $\Delta {\bi r}_i (t_0, t_1)$. 
}
\end{figure} 

\begin{figure}[t]
\includegraphics[width=1\linewidth]{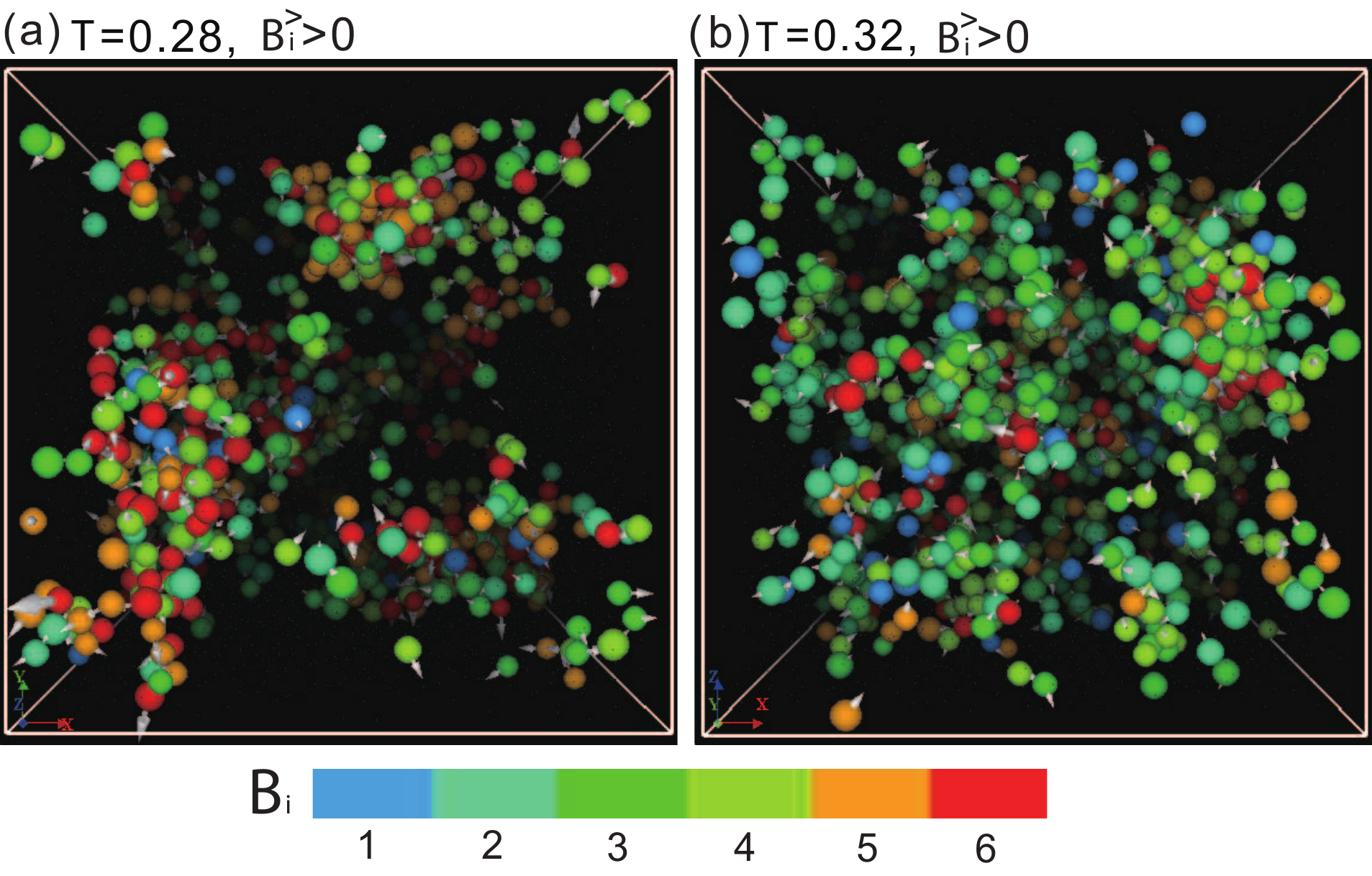}
\caption{(Color online) $\bi{BL}$ particles 
 (those with ${\cal B}_i^>(t_0,t_1)>0$) 
for (a) $T=0.28$  and  $t=t_1-t_0= 450$ and for  
(b) $T=0.32$ and $t=t_1-t_0=50$. The corresponding snapshot 
for $T=0.24$ is given in Fig.1b. 
Here, $t=t_1-t_0$ is determined from 
$\phi_b^>(t)=0.1$. 
The particle colors represent  ${\cal B}_i$
according to the color bar.The strings become ill-defined with increasing $T$. 
Arrows represents $\Delta {\bi r}_i (t_0, t_1)$. 
}
\end{figure}

In  Fig.1a, we display only the $\bi B$ particles (about 600) 
at  $t=t_1-t_0=10^4\cong 10^{-1}\tau_\alpha$ 
for  $T=0.24$, where 
the particle colors represent ${\cal B}_i= {\cal B}_i(t_0,t_1)$. 
We can see that   many particles with 
${\cal B}_i=1$ surround  those  
with ${\cal B}_i \ge 2$. Here,  the latter particles  
have undergone  stringlike  motions 
with  large displacements ($\gs 1$).  Thus, 
in  Fig.1b, we depict only the $\bi{ BL}$  particles 
with $\Delta r_i>0.8$.  Here, the  average  of $\Delta r_{i}$ 
over the $\bi{ BL}$ is 1.29, while  that over 
the $\bi{ BS}$ particles is  0.152 (see Fig.4 below). 
We also notice that a majority of 
the $\bi{BL}$ particles (74$\%$ here)  
belong to the first (small)  species (see Table 1 below). 
Thus, for  this time elapse of $10^4$, 
the  $\bi{BL}$ and  $\bi{BS}$  particles 
mostly form strings and their  
neighbors, respectively.

 At $T=0.24$,  a large fraction  
 of the $\bi{BS}$ particles surrounding  strings 
   transiently moved   
over relatively large distances ($>0.5$)   
at the string formation but eventually 
 returned  to their original positions (not shown in Fig.1b).  
Though rather rare, we also observed that 
some  particles largely moved  
without string  formation and returned  
to their original positions after some time (say, 400). 
These  {\it reversible jumps}  
were observed  by   Vollmayr-Lee \cite{Voll}.

However,   with increasing $T$, the  difference between 
the  $\bi{BL}$ and  $\bi{BS}$  particles 
becomes less distinct. 
In Fig.2, we elucidate 
 how the strings are changed   at  $T=0.28$ and 0.32,   
where the number fraction of $\bi{BL}$ particles $\phi_b^>$ 
is 0.1 (see Eq.(3.10) below for its definition). As compared to 
the snapshot at  $T=0.24$ in Fig.1b,  they are  shorter and more expanded 
with smaller ${\cal B}_i$, so they 
should rather be treated  as clusters. 
 This result  is consistent with the previous 
calculation of the distributions of 
 string lengths and cluster sizes  for various $T$ 
by Donati {\it et al}\cite{Kob,Glo}. 
With increasing $T$, 
we also notice that the composition of the large particles 
 increases  among the 
$\bi{BL}$ particles,  leading to larger 
volume fractions of the $\bi{BL}$ particles even at fixed $\phi_b^>=0.1$  
(see Table 1 below).

\begin{figure}[t]
\includegraphics[width=0.950\linewidth]{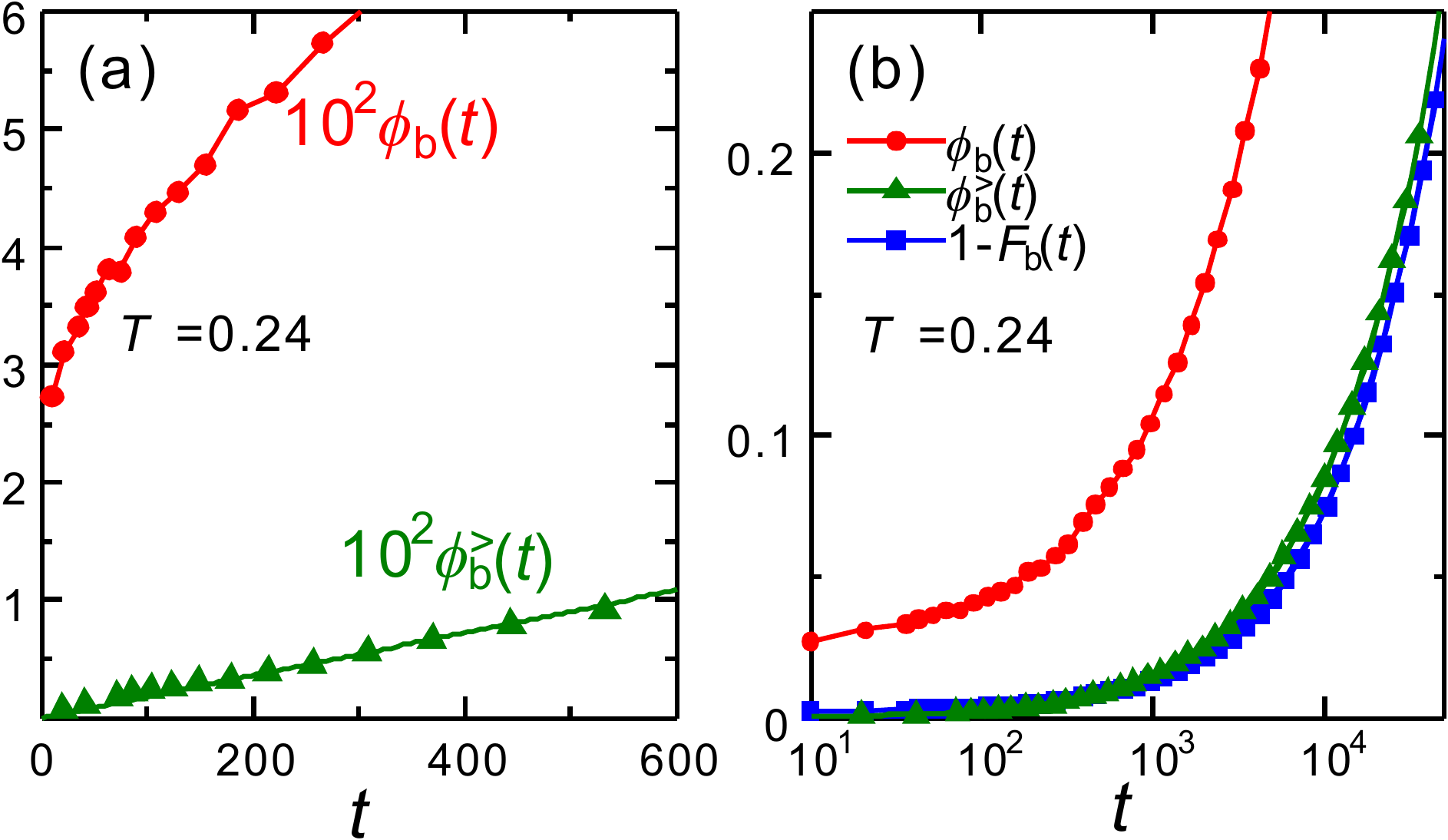}
\caption{(Color online) Fractions  of the $\bi B$ and $\bi{BL}$ 
particles, $\phi_b(t)$ 
and $\phi_b^>(t)$,   respectively, for $T=0.24$. They are shown 
for $t<800$ (left) and  on long times (right). 
Also displayed  is the fraction of 
surviving bonds $1-F_b(t)$ nearly coinciding  with $\phi_b^>(t)$ 
on long timescales (right).  }
\end{figure}

\begin{figure}[t]
\includegraphics[width=1\linewidth]{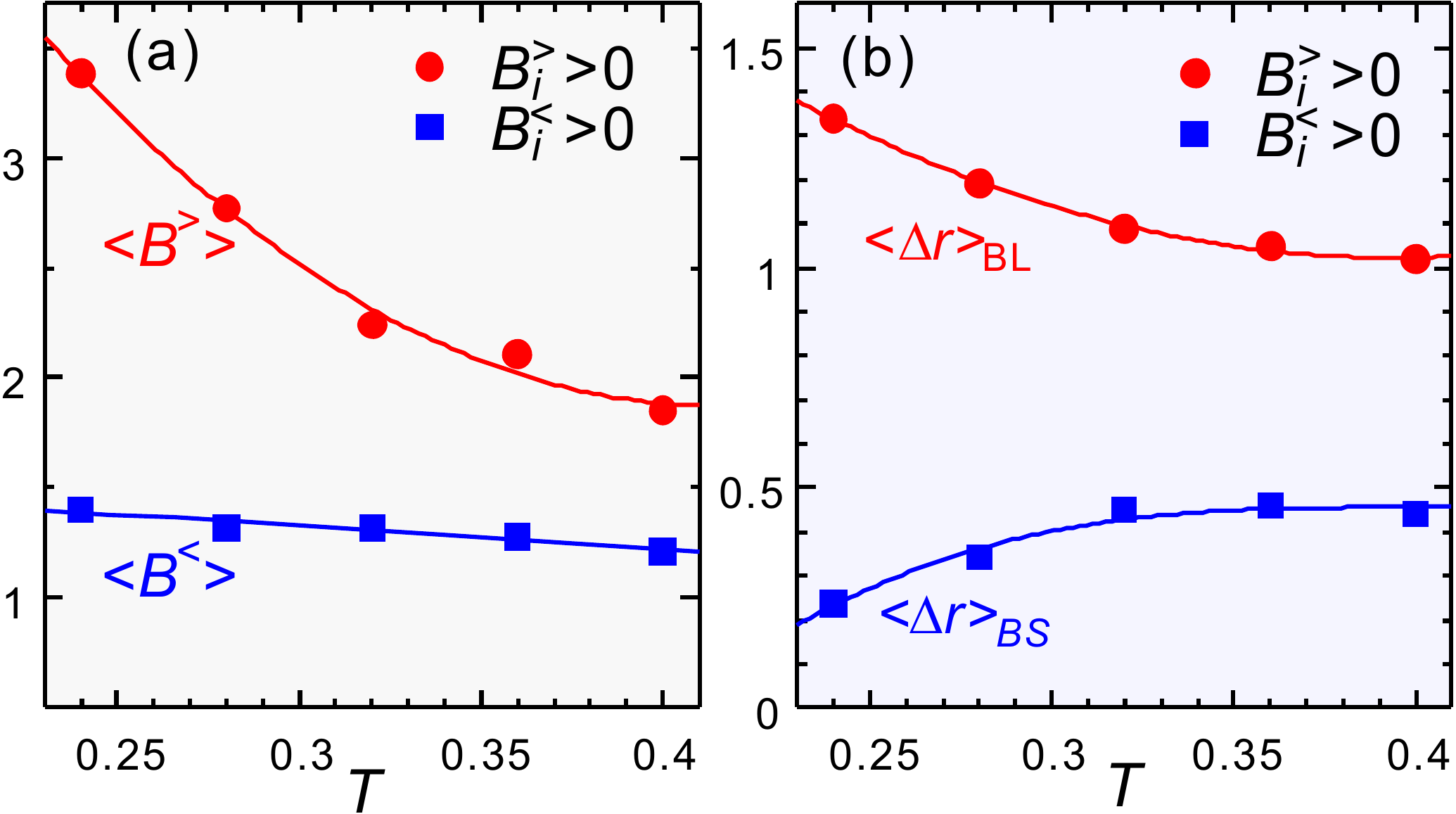}
\caption{(Color online) (a)  Averages    $\av{{\cal B}^>}$  
and  $\av{{\cal B}^<}$ in Eqs.(3.13) and (3.14) vs $T$. 
(b)  Averages    $\av{{\Delta r}_{BL}}$  
and   $\av{{\Delta r}_{BS}}$    in Eqs.(3.15) and (3.16) vs $T$. 
Here,   $\phi_b^>(t)=0.1$, from which 
 $t=t_1-t_0$ is determined for each $T$ as in  Table 1. 
}
\end{figure}


\begin{table}
\caption{Data  for five temperatures, where 
$t$,  $\phi_b(t)$,  $c_b^>(t)$,  and $c_b^<(t)$ 
are those  at $\phi_b^>(t)=0.1$. 
See  Fig.4 also. Here, 
 $\phi_b(t)=\phi_b^>(t)+\phi_b^<(t) $ 
is the fraction of the ${\bi B}$ particles, while 
 $c_b^>(t)$ and $c_b^<(t)$  are the compositions 
of the larger  species among  the $\bi{BL}$ and $\bi{BS}$ particles, 
respectively. 
 }
\begin{tabular}{|c||c|c||c|c|c|c|} 
\hline
$T$  &$\tau_\alpha$&$\tau_b$&$t$  &  $\phi_b(t)$ &$c_b^>(t)$  
$c_b^<(t)$\\
\hline
0.24 &109000&270000&10000 &  0.38 &0.247 &0.547 \\
\hline
0.28&1170&103000&450& 0.46&0.309 &0.496 \\
\hline
0.32&41.0&749&50& 0.59&0.329&0.492   \\
\hline
0.36&11.4&267& 20& 0.61&0.339&0.481 \\
\hline 
0.40&5.91&141 & 10& 0.59&0.330 &0.463 \\
\hline
\end{tabular}
\end{table}

\subsection{Differences between  $\bi{ BL}$ and 
$\bi{BS}$ particles }

Let us consider  the  numbers of 
the $\bi B$ and $\bi{BL}$ particles in the system.  The average  number 
fractions of these  particles are given by   
\bea 
\phi_b(t)&=&\frac{1}{N} \sum_j \av{\theta({\cal B}_j(t_0,t_1))}, \\ 
\phi_b^>(t)&=& \frac{1}{N} \sum_j \av{\theta({\cal B}_j^>(t_0,t_1))},
\ena 
where   $t=t_1-t_0$.   
The  $\phi_b(t)$ is the fraction of the particles with broken bonds, 
while $\phi_b^>(t)$ is 
that of the particles with broken bonds and large displacements. 
We may call $\phi_b^>(t)$ the string fraction also.  
  In our previous paper\cite{SKO}, 
we have introduced a  bond-preserving time $\tau_{\rm bp}$ by 
\be 
1- \phi_b(\tau_{\rm bp})=e^{-1},
\en 
which is the  timescale  of each particle 
to have a broken bond.  For  $T=0.24$, we have 
$\tau_b= 2.7\times 10^5= 2.7\tau_\alpha$ 
and  $\tau_{\rm bp}=3.2\times 10^4= 0.32 \tau_\alpha$.   
Note that 
$\tau_b$ is longer than $\tau_{\rm bp}$ by one order of magnitude 
due to the large coordination number in 3D.

 In Fig.3, $\phi_b(t)$ and $\phi_b^>(t)$ are plotted 
on short and long timescales.  Salient features are as follows. 
(i) First, on a 
 microscopic timescale ($t \gs  4$),   $\phi_b(t)$ quickly approaches 
 a small number$\sim 0.03$  
due to the thermal motions. (In 2D, 
we have obtained an algebraic growth, 
$\phi_b(t)\sim t^{0.6}$,  for small $t$, 
however \cite{SKO}.)  
(ii) Second, in the early stage ($t \ls 600$),  
$\phi_b^>(t)$ grows linearly  as  
\be 
\phi_b^>(t) \cong \tau_{\rm st}^{-1} t .
\en 
The coefficient 
$\tau_{\rm st}^{-1}$ is the average frequency 
of  rare jump motions per  particle 
in the early stage  $t \ll \tau_{\rm st}$ (where $\phi_b^>(t) \ll 1$). 
 Therefore, it is analogous to the nucleation rate in 
metastable systems \cite{Onukibook}. 
In our case,  $\tau_{\rm st}$ is of order 
$\tau_b$ in Eq.(3.4).  See Fig.5d for the $T$ dependence of  $\tau_{\rm st}$.  
(iii) Third,   in the whole time range in Fig.3, $\phi_b(t)$ 
is much larger than $ \phi_b^>(t)$.   
For $t \gg 1$, this is    because 
 several   broken bonds are produced around a string as in Fig.1a. 
 In Fig.3b, the ratio  $\phi_b(t)/\phi_b^>(t)$ is of order 5 
for $t\gs 10^3$. 
In addition,   $\phi_b^>(t)$  
is close to the broken bond fraction  
 $1-F_b(t)$, where $F_b(t)$ has appeared   in Eq.(3.4). 
 They are very close  for $T \ls 0.3$ in the present case.  
 (iv) Fourth, at long times  $t\gs 1000$, we have 
$\phi_b^>(t)\sim t^{0.76}$. 

In Fig.4, we  show  how $\bi{BL}$ and $\bi{BS}$ 
paricles behave  differently.  Displayed in the left  are 
the averages  of   ${\cal B}_i(t_0,t_1)$ 
among  the $\bi{ BL}$ and $\bi {BS}$ particles. They are written as  
\bea 
&&\av{{\cal B}^>}=\frac{1}{\phi_b^>N}{ \sum_i \av{{\cal B}_i^>(t_0,t_1)}}, \\
&&\av{{\cal B}^<}= \frac{1}{\phi_b^<N}{ \sum_i \av{{\cal B}_i^<(t_0,t_1)}}, 
\ena 
where  $\phi_b^<=\phi_b-\phi_b^>$ 
is the average fraction of the $\bi{BS}$ particles. 
The average broken-bond number  
is 3.4 for  the $\bi{ BL}$ particles 
and   is 1.4  for  the $\bi {BS}$ particles  at 
$\phi_b^>(t)=0.1$  for $T=0.24$. 
Displayed in the right  are the averages of the displacements 
$\Delta r_i$ in Eq.(3.6) among  the $\bi{ BL}$ and $\bi {BS}$ particles, 
\bea 
&&
\av{\Delta r}_{BL}= \frac{1}{\phi_b^>N}
\sum_{i }\av{ \Delta r_i \theta({\cal B}_i^>)}, \\ 
&&
\av{\Delta r}_{BS}= \frac{1}{\phi_b^<N}
\sum_{i }\av{ \Delta r_i \theta({\cal B}_i^<)}, 
\ena  
where  $\Delta r_i$,  ${\cal B}_i^>$, 
 and  ${\cal B}_i^<$ are  abbreviations 
of $\Delta r_i(t_0,t_1)$,  ${\cal B}_i^>(t_0,t_1)$, 
  and ${\cal B}_i^<(t_0,t_1)$, respectively.  
Remarkably, $\av{\Delta r}_{BL}$ and 
 $\av{\Delta r}_{BS}$ are distinctly separated 
and $\ell_{\rm m}$ in Eqs.(3.7) and (3.8) 
has bee chosen to be 0.8 between them. 
Table 1 presents more  data, 
where $c_b^>(t)$ and $c_b^<(t)$  are the compositions 
of the larger species among the $\bi{BL}$ and $\bi{BS}$ particles, 
respectively. The former is  the composition  within strings 
and  is only 0.25 at $T=0.24$.  

Thus   the motions of the 
$\bi{BL}$ and $\bi{BS}$ particles 
(those composing strings and those surrounding strings) 
are increasingly  different with lowering  $T$.  
This should be  a characteristic feature of fragile 
glass-forming liquids\cite{Angel,Ediger,Sastry}, where  
the role of the thermal fluctuations 
in the configuration changes 
 crosses over  with varying $T$.

 \begin{figure}[t]
\includegraphics[width=0.950\linewidth]{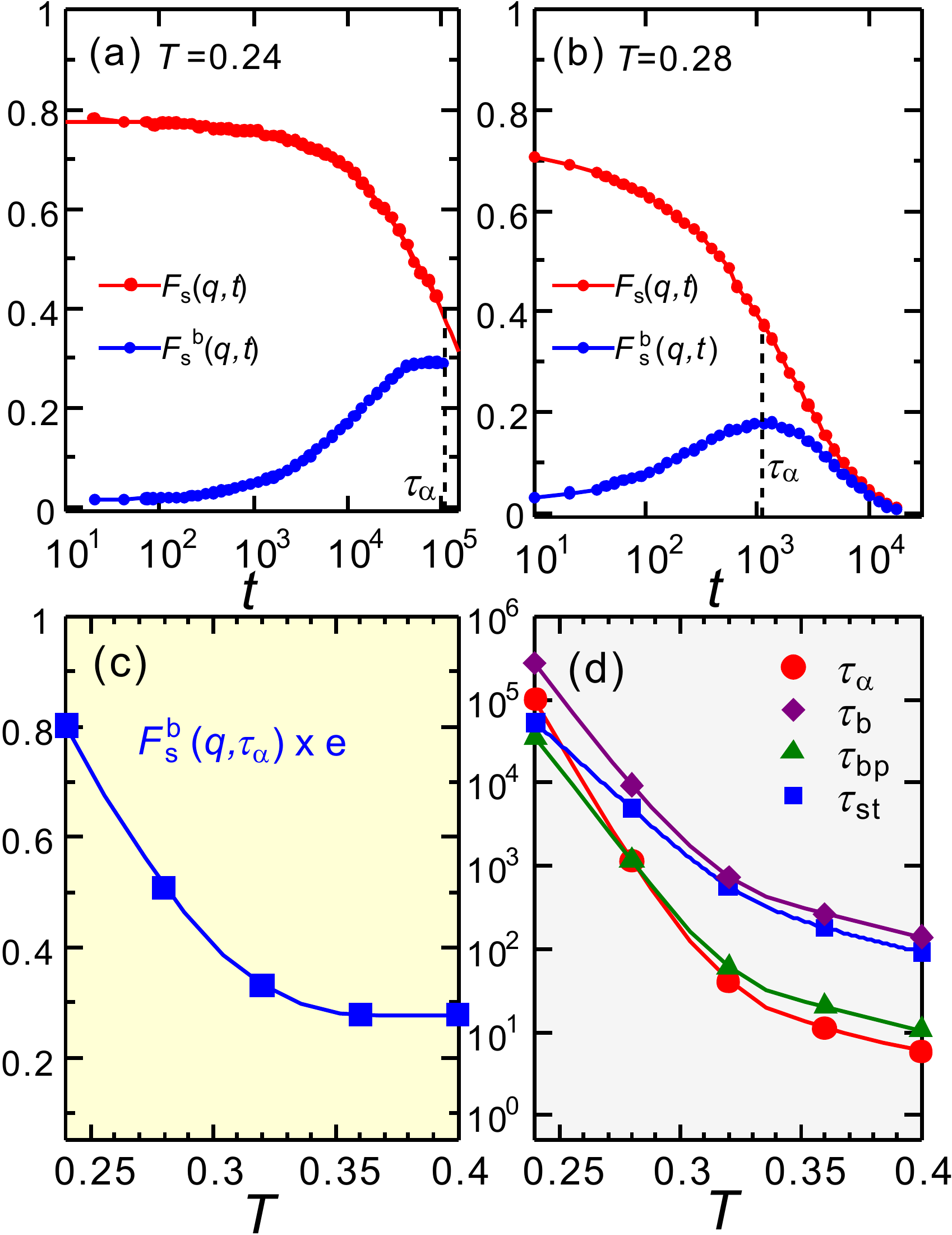}
\caption{ (Color online) Top: $F_s(q,t)$ in Eq.(2.4)  and  $F_s^b(q,t)$ in Eq.(3.17) 
for (a) $T=0.24$ and (b) $T=0.28$. 
Bottom:  $F_s^b(q,\tau_\alpha)$ 
vs $T$ in (c) and three characteristic  times $\tau_\alpha$ in Eq.(2.5), 
$\tau_{\rm bp}$ in Eq.(3.11),  $\tau_{\rm st}$ in Eq.(3.12), 
and $\tau_{b}$ in Eq.(3.4) in (d).  
Here,  $\tau_\alpha$ 
grows more steeply than the others at 
 low $T$. }
\end{figure}

 \subsection{Steep growth of $\tau_\alpha$ at low temperatures}

The physical meaning of  $\tau_\alpha$  is not straightforward. 
 First, $F_s(q,t)$ in Eq.(2.4) relaxes to a plateau 
value rapidly due to the thermal motions, 
which is about 0.8 for $T=0.24$   in Fig.5a 
and about 0.7 for $T=0.28$ in Fig.5b. 
 Second,  with lowering $T$,  
the motions of the $\bi{BS}$ particles (those surrounding strings) 
become  smaller  as in  Fig.4b, 
so their contribution   
to $F_s(q,t)$ becomes long-lived.

To examine  the second feature in more detail, 
we define the  conditional self-time-correlation function, 
\be 
F_s^b(q,t)= \frac{1}{N}
\sum_{ j} \av{  \theta({\cal{ B}}_j ) \exp[i{\bi q}\cdot (\Delta{\bi r}_j]},
\en  
where      ${\cal{ B}}_j$ and $\Delta {\bi  r}_j$  are abbreviations of 
${\cal{ B}}_j(t_0,t_1)$ and $\Delta {\bi r}_j(t_0,t_1)$, respectively, 
  and  only the $\bi B$ particles are picked up in the summation. 
In this function,  
 the contribution  from the $\bi{BL}$ particles  is less than 
$1\%$  at any $T$ (so ${\cal{ B}}_j$ may be replaced 
by ${\cal{ B}}_j^<$ in the 
right hand side of Eq.(3.17)). 
In (a) and (b) of Fig.5, $F_s(q,t)$ and $F_s^b(q,t)$ 
are compared at   $T=0.24$ and 0.28. 
In (c),  the $T$ dependence of $F_s^b(q,\tau_\alpha)$ is shown at  
$t=\tau_\alpha$. It 
is about $30\%$ of  $F_s(q,\tau_\alpha)=e^{-1}$ 
for  relatively high $T$,  but it increases  up to $80\%$ 
for  $T=0.24$. This dramatic change serves to slow down  the 
relaxation  of $F_s(q,t)$  at low $T$.  
In (d), we display 
 $\tau_\alpha$, 
$\tau_{\rm bp}$,  $\tau_{\rm st}$, 
and $\tau_{b}$,  where 
$\tau_\alpha$ indeed grows most  steeply 
with lowering $T$. Due to this crossover, 
we have 
 $\tau_b/\tau_\alpha \sim 20$ for 
$\tau_\alpha \ls 10^4$ 
but  $\tau_b/\tau_\alpha \sim  2.7$ for  
$\tau_\alpha \sim 10^5$ from Table 1.

\section{Diffusion } 
\setcounter{equation}{0}

\subsection{Contributions to the mean square displacement} 

It is  widely believed that the diffusion 
of a tagged or test particle  in supercooled and glassy systems  is 
caused by its jump or escape motions    from 
temporal cages 
\cite{Binder,Silescu,Ediger,Harrowell,Wa,yo-diffusion,Rei,Sch,Szamel}. At  $T=0.24$,  we  show that the diffusion occurs as  
activation processes, where 
only the $\bi{ BL}$ particles can diffuse over long distances.

\begin{figure}[t]
\includegraphics[width=0.950\linewidth]{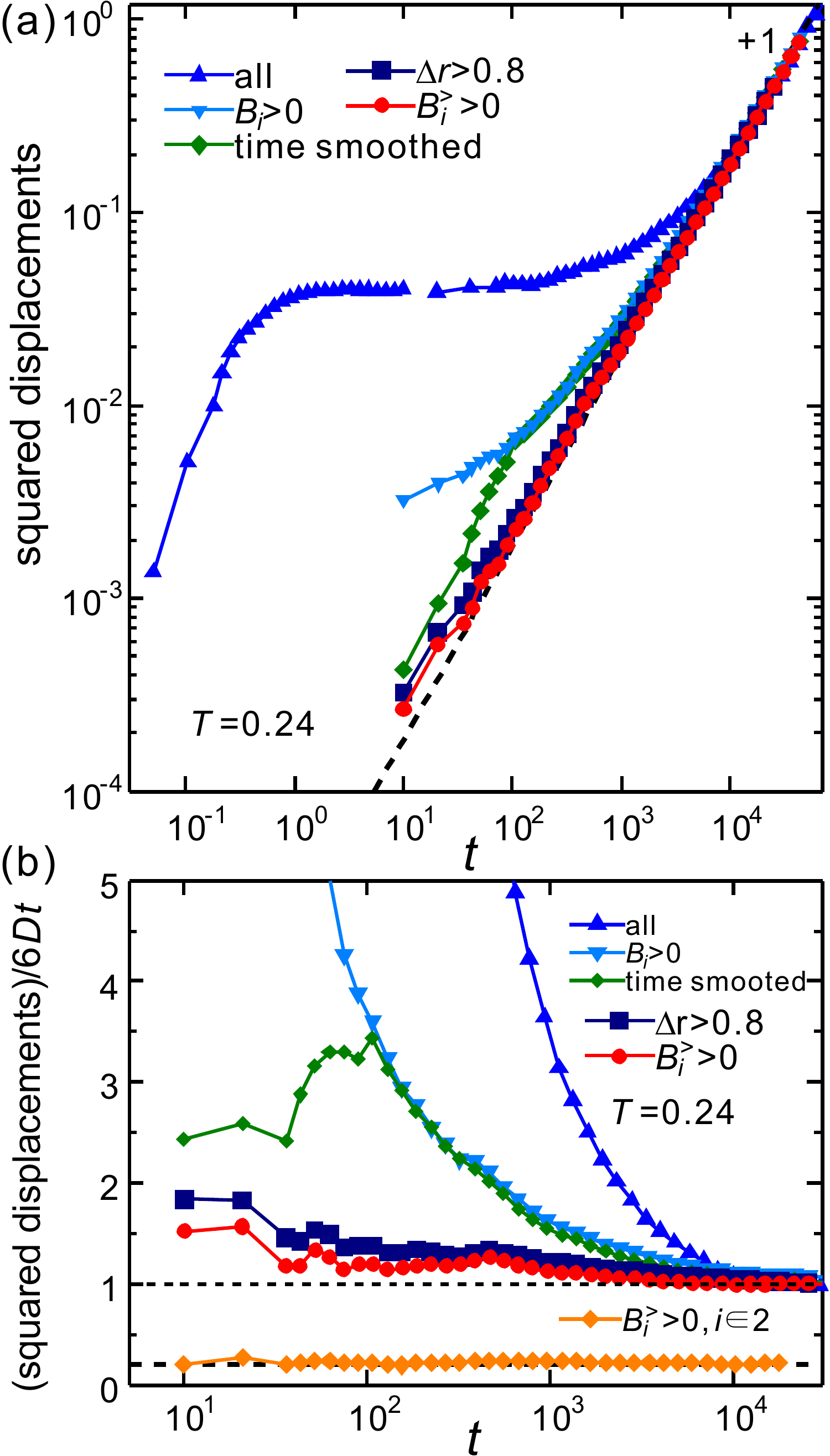}
\caption{(Color online) Contributions to the mean square displacement 
  for  $T=0.24$.   
 (a)  $M(t)$ 
from all the particles  (top curve),  
 $M_B(t)$ from $\bi B$ particles, 
$M_{B}^>(t)$ from $\bi{ BL}$ particles,  
 $M^>(t)$ from  those  with $\Delta r_i>0.8$, and
 ${\bar M}(t)$ for  time-smoothed positions on a logarithmic scale. 
See  Eqs.(4.4)-(4.7).  
(b) Top five curves represent 
contributions  
divided by $6Dt$ on  a semi-logarithmic scale, 
 tending to unity with increasing $t$. The approach 
is very  rapid for $M^>(t)/6Dt$ and $M_B^>(t)/6Dt$. 
Shown also is the contribution divided by $6Dt$ 
from $\bi{ BL}$ particles 
belonging to the second species (bottom line), 
which is very close  to  
$D_2/(D_1+D_2) \cong 0.2$ for $t \gs 10$.   
}
\end{figure}

\begin{figure}[t]
\includegraphics[width=1\linewidth]{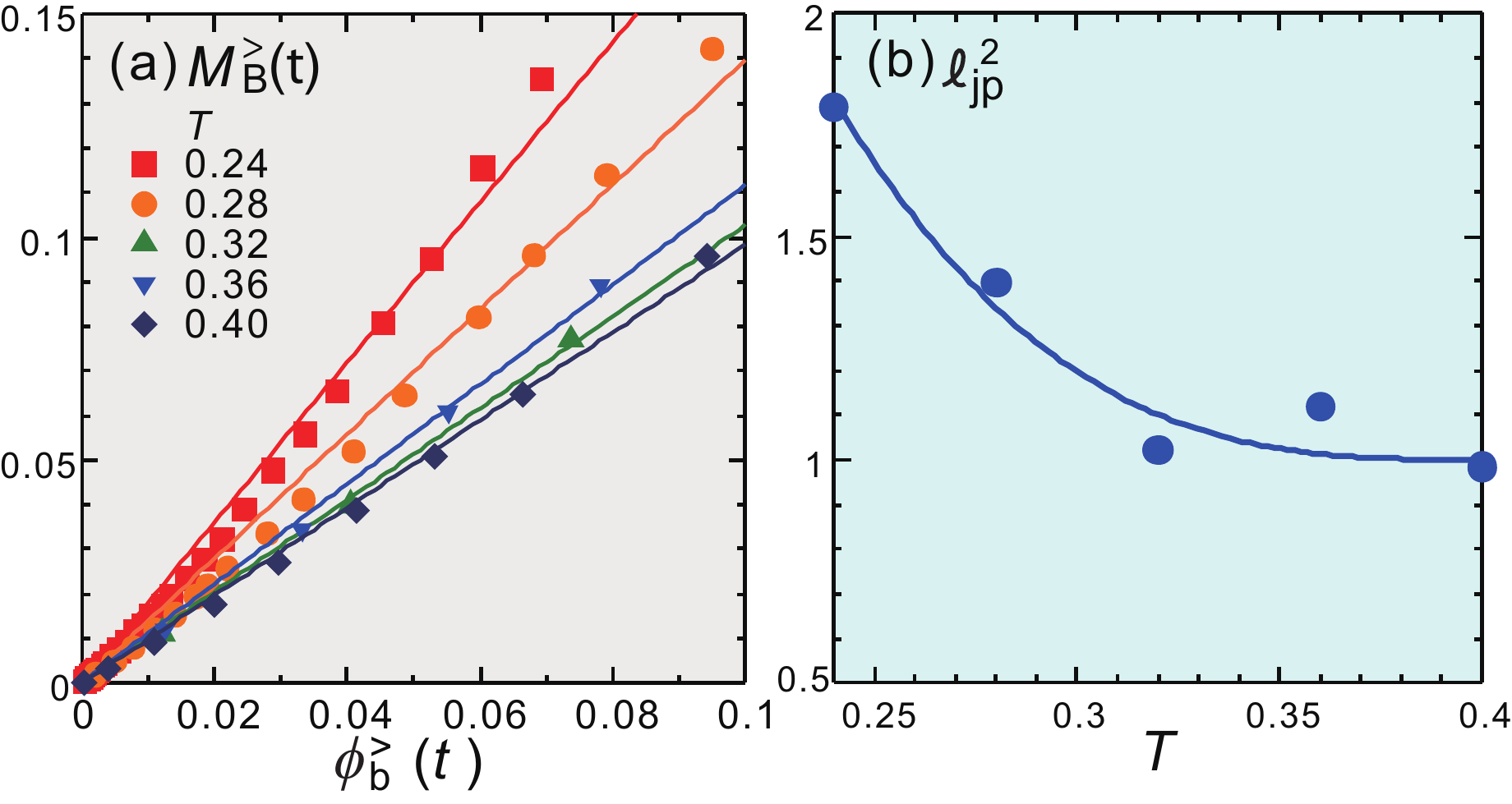}
\caption{ (Color online) (a)  $M_B^>(t)$ vs $\phi_b^>(t)$ for various $T$,  
 where both grow linearly in time in the early stage. 
(b) Ratio $\ell_{\rm jp}^2= M_B^>(t)/\phi_b^>(t)$ vs $T$, where   
$\ell_{\rm jp}$ is a characteristic jump length of a particle. }
\end{figure}

The mean-square displacement  $M(t) $ is  written as  
\be 
M(t)= \av{|\Delta r (t)|^2}= 
\frac{1}{N}\sum_i \av{(\Delta {r}_i(t_0,t_1))^2}
\en  
where $\Delta { r}_i(t_0,t_1)$ 
 is the  displacement length   in Eq.(3.6) 
in time interval with  width $t= t_1-t_0$. As is well known, 
 $M(t)$ exhibits the ballistic behavior 
($\propto t^2$) in the very early stage ($t \ls 1$), the   
 plateau behavior $M(t) \cong M_{\rm p}$ in an  intermediate time range, 
and the  linear growth  $M(t) \cong 6 Dt$ 
 in the late stage. 
In $M(t)$, the diffusion regime starts  after  
dominance of the jump motions over the thermal vibrational motions. 
 Since  we take the average over all the particles 
in Eq.(4.1),  $D$ in this paper is the following mean value,
\be 
D= (1-c)D_1+ cD_2,
\en 
where  $D_1$ and $D_2$ are the diffusion constants of the 
first and second species, respectively, and $c$ is the composition.

In Fig.6a, the top curve represents    $M(t)$, which exhibits 
its   typical behaviors mentioned above 
with  $M_{\rm p}\cong  0.036$. 
Other three curves represent  the following contributions, 
\bea 
M_{B}(t)&=&
\frac{1}{N}\sum_{i}
 \av{(\Delta { r}_i)^2\theta({\cal B}_i)} ,\\ 
M_{B}^>(t)&=&
\frac{1}{N}\sum_{i}
 \av{(\Delta { r}_i)^2\theta({\cal B}_i^>)} ,\\
 M^>(t)&=& \frac{1}{N} \sum_{i}
\av{(\Delta { r}_i)^2\theta(\Delta r_i-\ell_{\rm m})},
\ena  
where  $\Delta { r}_i$, ${\cal{ B}}_i$, 
 and ${\cal{ B}}_i^>$  are abbreviations of 
$\Delta { r}_i(t_0,t_1)$,  ${\cal{ B}}_i(t_0,t_1)$,  
and  ${\cal{ B}}_i^>(t_0,t_1)$, respectively, and $\ell_{\rm m}=0.8$. 
Here, owing to  the step function $\theta(\cdots)$,  
 $M_B(t)$ arises from the $\bi B$ particles,  
  $M_B^>(t)$ from the $\bi{ BL}$ particles, 
and $M^>(t)$  from the particles with  $\Delta r_i> 0.8$. 
In addition, defining the smoothed 
position 
\be 
{\bar{\bi r}}_i (t)=  \frac{1}{\Delta t}
\int^t_{t-\Delta t}dt' {\bi r}_i(t') 
\en 
 with $\Delta t=100$, 
we calculate the time-smoothed mean square displacement,  
\be 
{\bar M}(t)=\frac{1}{N}\sum_{i}
 \av{|{\bar{\bi  r}}_i(t_0+t)- {\bar{\bi  r}}_i(t_0)|^2}.   
\en 
In Fig.6, we can see the sequence  
${ M}_B(t)>{\bar M}(t)>{M}^>(t)>M_B^>(t)$. 
To see the approach to the diffusion behavior, 
 in Fig.6b, we 
 display  $M(t)/6Dt$,  $M_{B}(t)/6Dt$,  ${\bar M}(t)/6Dt$, 
$M^>(t)/6Dt$, 
and ${ M}_B^>(t)/6Dt$ with $D=0.596\times 10^{-5}$.  
The latter four  quantities approach 
unity more rapidly than $M(t)/6Dt$. 
In particular,   the jump contributions 
$M^>(t)$ and   $M_B(t)^>$ become very close to 
$6Dt$   from the very   early stage 
$t\gs 50 \sim 5\times 10^{-4}\tau_\alpha$.

To be precise, we should calculate $D_1$ and $D_2$ separately. 
To this end, we also consider 
 the contribution from the  $\bi{BL}$ particles 
belonging to the second species,
\be 
M_{B2}^>(t)=
\frac{1}{N}\sum_{i\in 2}
 \av{(\Delta { r}_i)^2\theta({\cal B}_i^>)} .
\en 
In Fig.6b, the ratio  $M_{B2}^>(t)/6Dt$ is plotted  
 as a   flat  line at the bottom.  
It is  equal to 0.21  
 for $t\gs 10\sim 10^{-4}\tau_\alpha$. This value should be 
equal to $cD_2/D = D_2/(D_1+D_2)$, 
 so we have    the 
diffusion constant ratio $D_2/D_1 \cong 0.27$ in this case. 
In accord with  this ratio, 
the composition of the large particles 
among the $\bi{BL}$  particles is    $c_b^>(t)\cong 0.25$ for $T=0.24$ 
  in Table 1.

Recall that 
the fraction of the $\bi{BL}$ particles $\phi_b^>(t)$  
 grows linearly in time in the early stage in Fig.3,  leading to 
the characteristic time $\tau_{\rm st}$  in Eq.(3.12).  
We may  define   a characteristic  jump length  $\ell_{\rm jp}$ 
from the following ratio,    
\be 
\ell_{\rm jp}^2= M_B^>(t)/ \phi_b^>(t). 
\en 
Now  $D$ is expressed as 
 \be 
D= \ell_{\rm jp}^2/6\tau_{\rm st}. 
\en 
In Fig.7a, we plot $\phi_b^>(t)$ vs $M_B^>(t)$ 
 to show their linear relationship. In  Fig.7b,  
we plot $\ell_{\rm jp}^2$ vs $T$, where 
$\ell_{\rm jp}$ is close to the particle diameter $\sigma_1$, 
 only weakly depending  on $T$. As a  natural result, 
 its   $T$-dependence  is   similar to that of 
the average displacement length 
$\av{r}_{BL}$  in Fig.4b. 
In our calculation of $\phi_b^>(t)$ and  $M_B^>(t)$, 
the lower bound of the jump motions $\ell_{\rm m}$ 
is set equal to 0.8 for any $T$ (see Subsec.IVD 
 for comments on this aspect).

Previously  \cite{Harrowell,yo-diffusion},   $M(t)$ 
was divided into  two contributions 
from appropriately defined mobile and immobile 
particles, where the former turned out to dominate over the 
latter  in the diffusion regime, in accord with the results in this 
section.

\subsection{Violation of the Stokes-Einstein relation}

In our model particle system, the violation of 
the Stokes-Einstein relation 
 is ascribed to the steep growth 
of $\tau_\alpha$ at low $T$, 
which has been discussed in Sec.IIIC. 
However, we might need to include other elements (such as the 
nonsphericity of the particles) to  explain 
the observed diffusion behavior in real molecular  
 systems   \cite{Silescu,Ediger,Glo}.

In  Fig.8a, 
we plot $D=(D_1+D_2)/2$, $D_1$,  and $D_2$ vs $T$. For $T\ge 0.28$ 
we calculated them from the usual mean square displacements.  
Here, $D_2/D_1\cong 0.27$ for  $T=0.24$ and $0.71$  for $T=0.40$. 
Thus, $D_2$ decreases more rapidly than $D_1$ 
with lowering   $T$, which is consistent with the 
$T$ dependence of the composition of the $\bi{BL}$ particles in Table 1. 
In Fig.8b, we plot the products $D\tau_\alpha$, 
$D_1\tau_\alpha$,  $D_2\tau_\alpha$,  ,   
and $D\tau_{b}$ vs  $T$.
The behavior of $D_1 \tau_\alpha$   is nearly the same 
as in the previous simulation\cite{yo-diffusion}. 
However,   $D\tau_{b}$ 
depends  on $T$ much more  weakly  than  
  $D\tau_\alpha$, where  
 $\tau_{b}$ is the life time of bonds in Eq.(3.4).  
From Eq.(4.10) and Fig.7b, we notice that the relation 
$\tau_{\rm st }\cong \tau_{b}/3$ well holds 
in the temperature range studied.  These results indicate 
 that the diffusion is  governed by  the activated dynamics, 
while the stress relaxation is more 
sensitive to the thermal fluctuations  
in fragile systems.

From Fig.8b,  the  violation of 
the Stokes-Einstein relation is weaker for a large  particle 
than for a small particle.  The participation of  
larger particles in the jump motions 
should become increasingly infrequent with lowering $T$, as suggested    
by  the previous experiments \cite{Silescu,Ediger}.

\begin{figure}[t]
\includegraphics[width=1\linewidth]{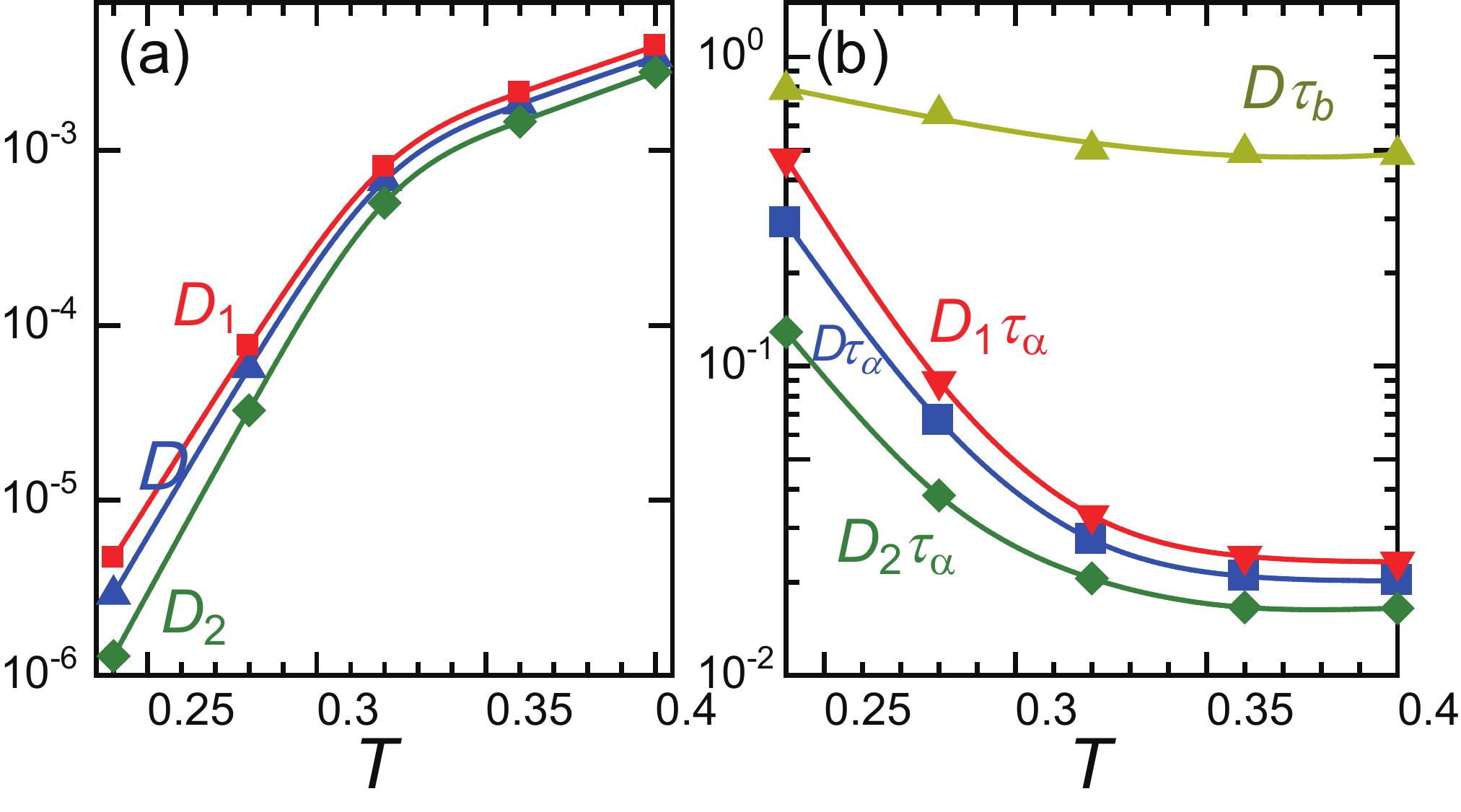}
\caption{(Color online) (a)  Diffusion constants $D_1$ for small particles, 
    $D_2$ for large particles,  and $D=(D_1+D_2)/2$ as functions of 
 $T$. (b) $D_1\tau_\alpha$,  $D_2\tau_\alpha$, and  
$D\tau_\alpha$, and   $D\tau_{b}$ as functions of  $T$. The curves of 
  $D_1\tau_\alpha$  and   $D_2\tau_\alpha$ 
indicate the violation of the Stokes-Einstein relation 
from $\eta/T \propto \tau_\alpha$, whereas $D\tau_b$ 
only weakly depends on $T$.   
}
\end{figure}

 \subsection{Van Hove self-correlation function } 
\begin{figure}[t]
\includegraphics[width=0.950\linewidth]{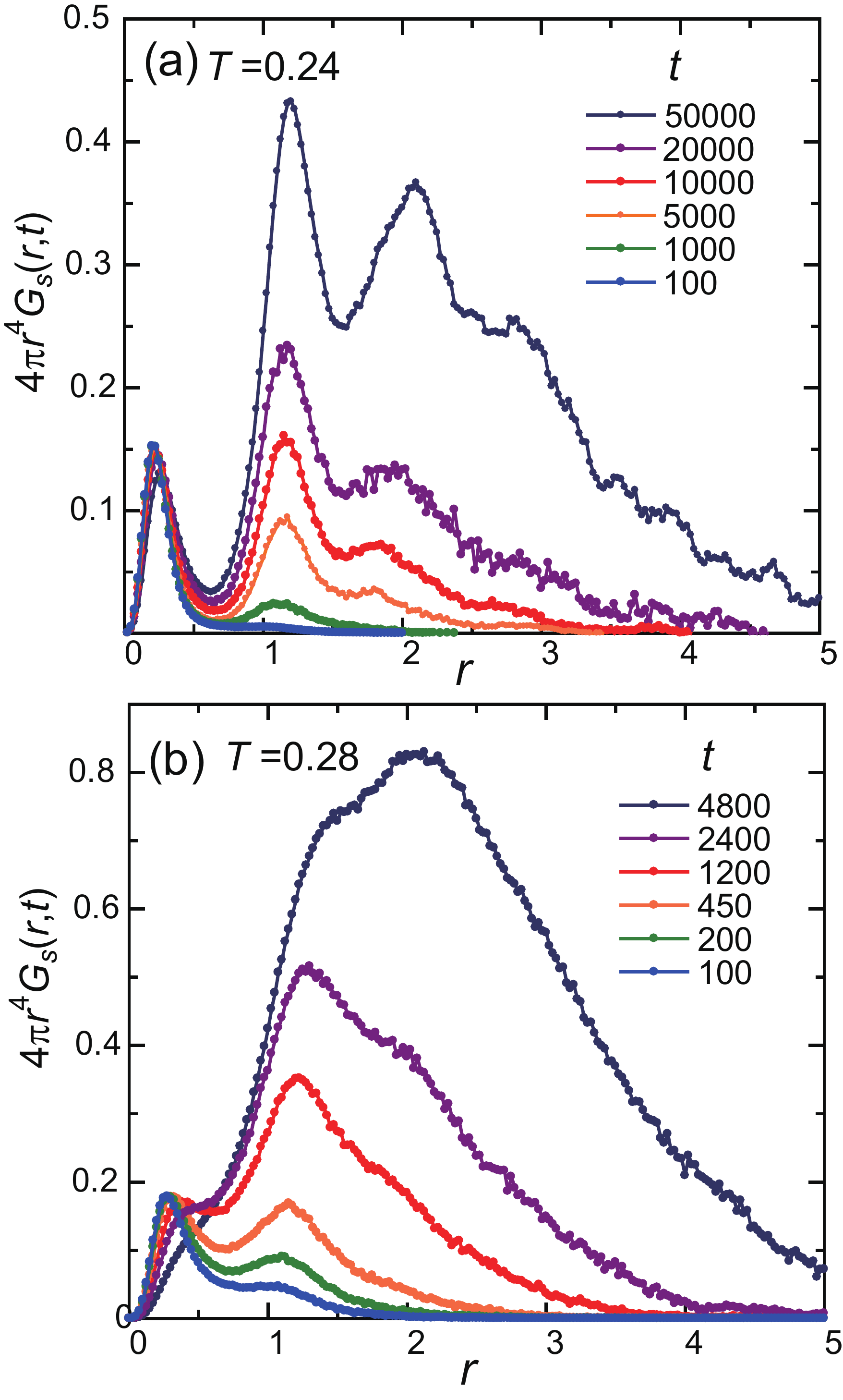}
\caption{(Color online) $4\pi r^4 G_s(r,t)$ vs $r$ at various $t$.  
(a) For  $T=0.24$, there is a deep minimum at $r= r_{\rm m} \sim 0.7$, 
which separates a nearly stationary  part 
for $r<r_{\rm m}$ and a  growing  part with multiple maxima  
for $r>r_{\rm m}$.  
(b) For $T= 0.28$, a minimum at $r= r_{\rm m} 
\sim 0.7$ increases in time with one maximum for 
$r>r_{\rm m}$.   A secondary  peak in the outer region 
$r>r_{\rm m}$  starts to appear as a shoulder.
}
\end{figure}

\begin{figure}
\includegraphics[width=0.980\linewidth]{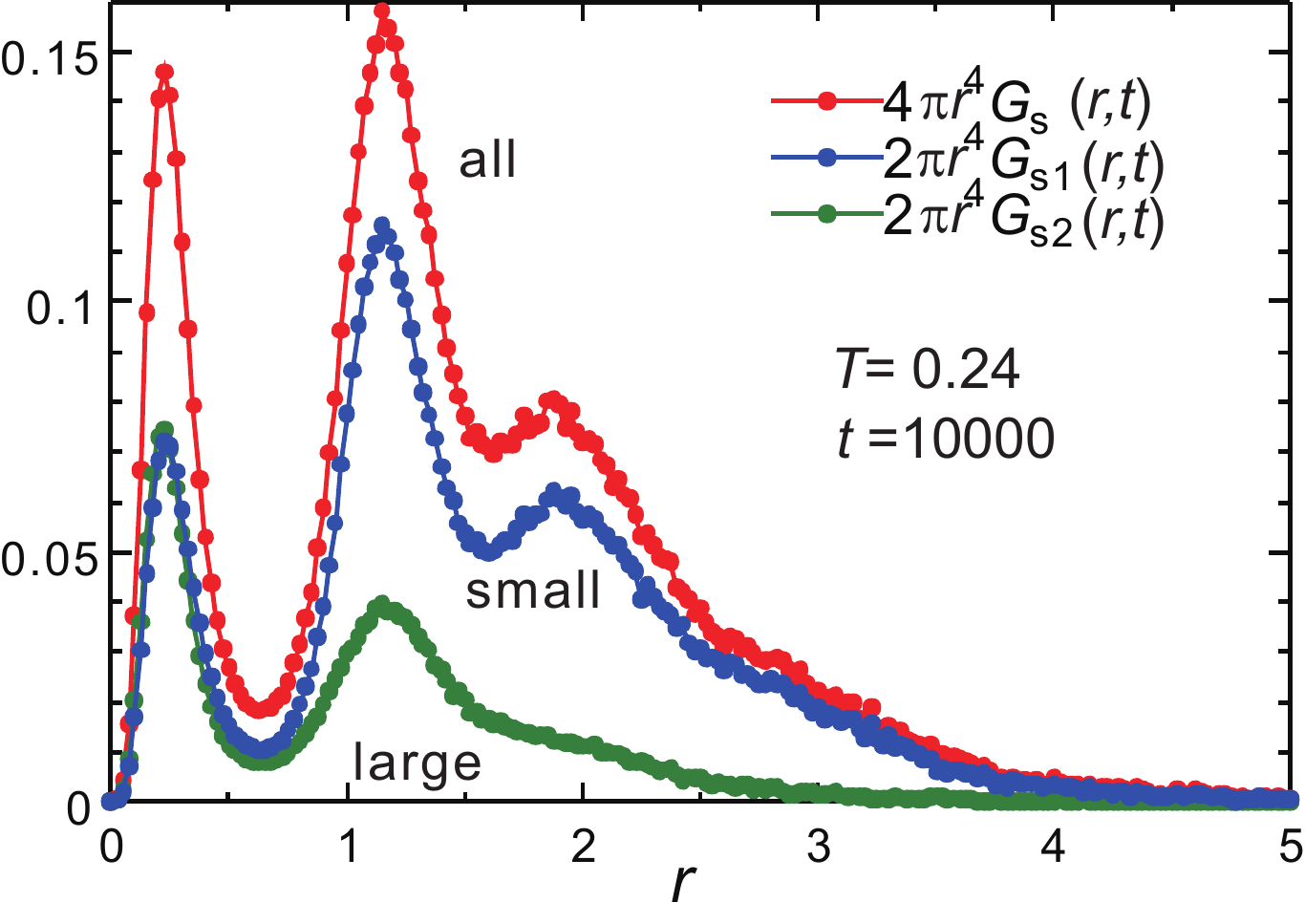}
\caption{(Color online) $4\pi r^4 G_{s}(r,t)$ 
is decomposed into the small particle 
part  $2\pi r^4 G_{s1}(r,t)$ 
and the large particle part 
$2\pi r^4 G_{s2}(r,t)$ from Eq.(4.12)
 at   $t=10^4$ for $T=0.24$. The minimum at 
$r=r_{\rm m}$ is common for the small and large particles.
}
\end{figure}

\begin{figure}
\includegraphics[width=0.980\linewidth]{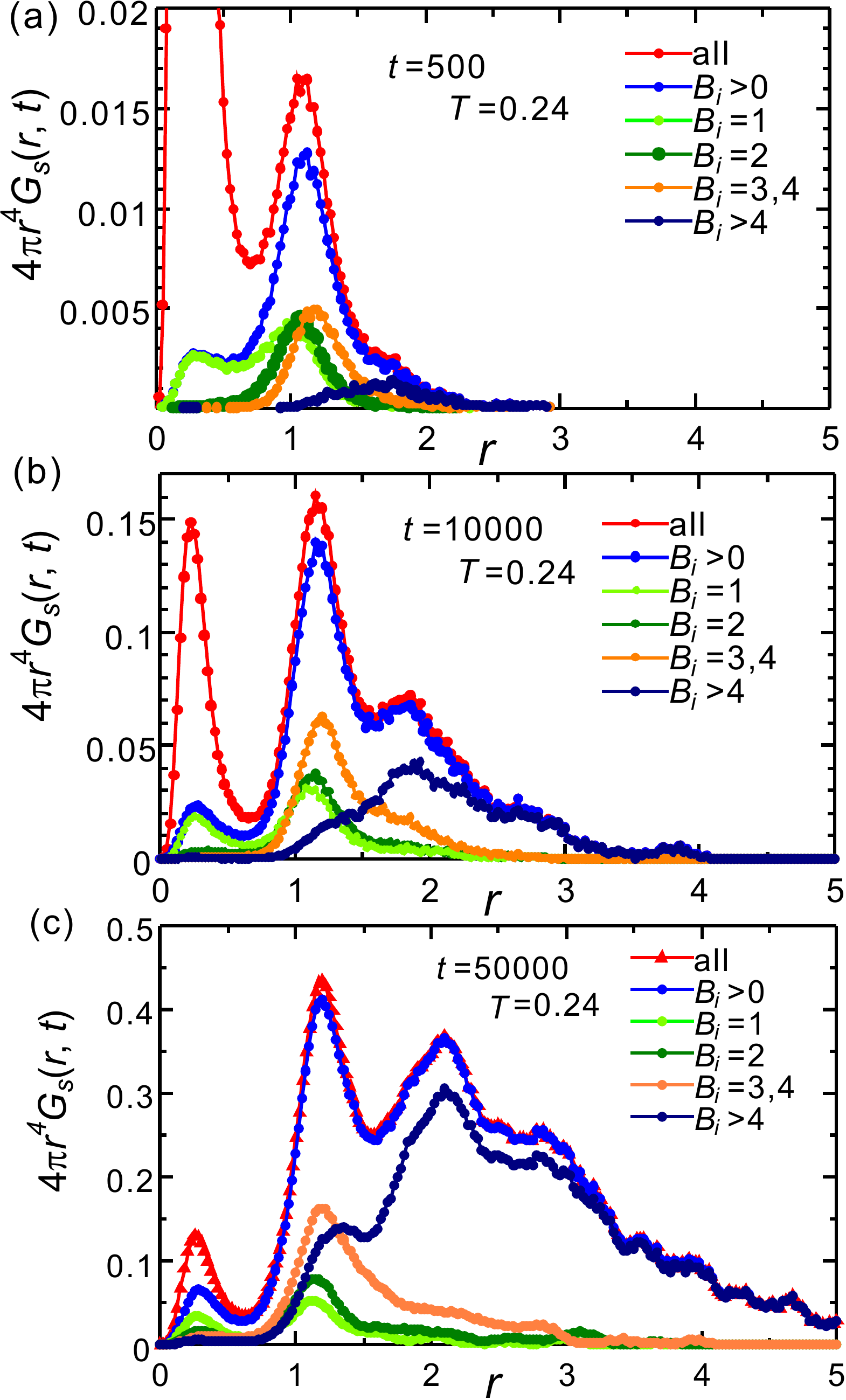}
\caption{(Color online) $4\pi r^4 G_s(r,t)$ vs $r$ at   (a) $t=500$, (b) 10000, 
and (c) 50000 for $T=0.24$. The contributions from the particles 
with ${\cal B}_i(t_0,t_0+t)=k$ are shown, which have $k$ broken bonds 
in time interval $[t_0,t_0+t]$. 
}
\end{figure} 

The time-correlation function 
 $F_s(q,t)$ in Eq.(2.4) has been used so far, though it   has been 
calculated only for $q=2\pi$. Its inverse Fourier 
transformation  $G_s(r,t)$, 
called the van Hove self-correlation function,  
exhibits thermal activation  behavior \cite{Sastry} as a function of $r=|{\bi r}|$ and $t$ 
as  $T$ is lowered. 

From Eq.(2.4) we obtain     
\be 
G_{s}(r,t)=\frac{1}{N}    
\sum_{j} \av{\delta (\Delta{\bi r}_j(t_0, t_0+t)-{\bi r}) } ,   
\en   
which is normalized as $4\pi \int_0^\infty dr r^2 G_s(r,t)=1$.   It  
represents the probability of finding a particle with displacement $\bi r$ 
after  time $t$.  In our definition,   the average of 
$\av{\delta (\Delta{\bi r}_j(t_0, t_0+t)-{\bi r}) }$  
is taken over  all the particles, so it is the mean 
of the van Hove self-correlation functions for the two species as 
\be 
G_{s}(r,t)=(1-c)G_{s1}(r,t)+c G_{s2}(r,t). 
\en 
Here,   $G_{s\alpha}(r,t)$ ($\alpha=1,2$) are 
the averages over the particles of the species $\alpha$ 
with  $N^{-1}\sum_j$  in Eq.(4.11) being replaced by  
$N_\alpha^{-1}\sum_{j\in \alpha}$.   
The mean square displacement $M(t)$ in Eq.(4.1) is expressed as 
\be
M(t) =  4\pi \int_0^\infty dr~ r^4 G_s(r,t). 
\en  
The  $M^>(t)$ in Eq.(4.5) follows 
if the lower bound   of the above integral 
is increased to 0.8.

In Fig.9, we display $4\pi r^4 G_s(r,t)$ vs 
$r$ at various  $t$. The area below its curve is equal to 
$M(t)$.  For $T=0.24$ in (a), it  
exhibits a deep minimum  
at  $r=r_{\rm m}\sim 0.7$. In the interior  $r<r_{\rm m}$ 
it is nearly stationary, 
while in the exterior  $r>r_{\rm m}$ it 
grows with  two or three maxima 
arising from multiple jumps. 
From Fig.6, the area of the outer region 
is nearly equal to $6Dt$ for each curve.  
For $T=0.28$ in (b), in contrast, 
a shoulder first appears  to grow into a peak 
in the exterior $r > r_{\rm m}$. 
The  minimum at $r = r_{\rm m}$ 
increases considerably  in time.

From Eq.(4.12),  $G_s(r,t)$ is equal to 
the mean $G_{s1}(r,t)/2+ G_{s2}(r,t)/2$ for $c=1/2$. 
We are also interested in the difference of the 
motions  of the small and large particles. Therefore, 
in Fig.10, we plot the three curves of 
$4\pi r^4 G_{s}(r,t)$,  $2\pi r^4 G_{s1}(r,t)$, 
and  $2\pi r^4 G_{s2}(r,t)$ at $t=10^4$ for $T=0.24$.  
All these curves  exhibit a deep minimum 
at the common length $r= r_{\rm m}$.  Because  $D_2/D_1=0.27$ here, 
the area of   $2\pi r^4 G_{s1}(r,t)$  
 in the exterior   $r> r_{\rm m}$ 
is four times larger than that 
of   $2\pi r^4 G_{s2}(r,t)$. 
A second peak corresponding to two jump motions 
can be seen for the small particle but 
not for the the large particle, although 
the large particles also undergo multiple jumps 
on long timescales.

In Fig.11, we furthermore 
divide $G_s(r,t)$ into the contributions from 
${\cal B}_i(t_0,t_0+t)=k$ with $k=0,1,\cdots$ for three times.
The particles with $k >4$ have undergone multiple jumps.  
In (a) at $t=500$, those with $k \le 4$ 
have undergone a single jump, while those with $k>4$ 
two jumps.  In (b) at $t=10^4$, the fraction of 
the particles with $k>4$ is increased. 
In (c) at $t=5\times 10^4$, the fraction of 
three jumps becomes noticeable yielding a small third peak. 
We can thus see how multiple stringlike jump motions 
give rise to these contributions.

Previously, for the Lennard-Jones potential, 
Sastry {\it et al.}\cite{Sastry} and  Wahnstr$\ddot{\rm o}$m 
\cite{Wa} found  secondary peaks in 
$  4\pi r^2 G_{s}(r,t)$ at low $T$. 
For hard-sphere binary mixtures 
  \cite{Rei,Sch,Szamel}, 
 the same behavior  was noticed 
for sufficiently large volume fractions of the 
particles.  In particular, Reichman {\it et al.}\cite{Rei} 
found appearance of two peaks in the outer region $r>r_{\rm m}$ 
indicating discrete step motions of hard spheres.

The  behaviors of $G_s(r,t)$ at $T=0.24$ 
can be expected generally  for glassy particle systems at low $T$. 
A particle in such systems is analogous to 
a Brownian particle slowly escaping from  
a cage represented by a potential $U(r)$. 
If the barrier height   $U_{\rm m}= 
U(r_{\rm m})$ at $r=r_{\rm m}$ is  much higher than $k_BT$,   
 the  probability  of staying around this barrier  is  
very small ($\propto  \exp[-U_{\rm m}/k_BT]$)  and 
the probability of escaping 
to  the outer region  is proportional to $t$. 
The  same behaviors  have indeed been found for  $G_s(r,t)$ 
at  $T=0.24$ in Fig.9a, where 
the minimum length $r_{\rm m}$ is analogous to 
the critical radius in metastable systems.

\subsection{Dependence on the cut-off length $\ell_{\rm m}$  }

In this paper,  the lower bound $\ell_{\rm m}$ of the 
jump motions  in Eqs.(3.7) and (3.8)  has been  set equal to 0.8.  
As remarked below Eq.(3.16), 
this length  is between the 
average displacement lengths $\av{\Delta r}_{BL}$  and 
 $\av{\Delta r}_{BL}$   for the  $\bi{BL}$  
and $\bi{BS}$ particles in Eqs.(3.15) and (3.16), where 
 $\av{\Delta r}_{BL}>1$  and 
 $\av{\Delta r}_{BL}<0.5$ in Fig.4b. 
We also note 
that   this   cut-off length 
is close  to the minimum 
distance  $r_{\rm m}\sim 0.7$ in 
$4\pi r^4 G_s(r,t)$ in  Fig.9a.  
In Fig.12, by setting $\ell_{\rm m}=0,7, 0.8$, and $0.9$, 
we plot the three corresponding curves of  $M_B^>(t)/6Dt$. 
They  surely tend to unity for $t\gs 10^3$, 
so   $M_B^>(t)$ is   insensitive to a small 
change of $\ell_{\rm m}$.

At  $T$ higher than   $ 0.3$, however, 
the cage life time 
becomes  shorter. As a result,  
 $\phi_B^>(t)$  in Eq.(3.10),   
$\tau_{\rm st}$  in Eq.(3.12), and $\ell_{\rm jp}$  in Eq.(4.9) 
 significantly depend 
on  $\ell_{\rm m}$, although 
we have set  $\ell_{\rm m}=0.8$ for any $T$.

\begin{figure}
\includegraphics[width=0.88\linewidth]{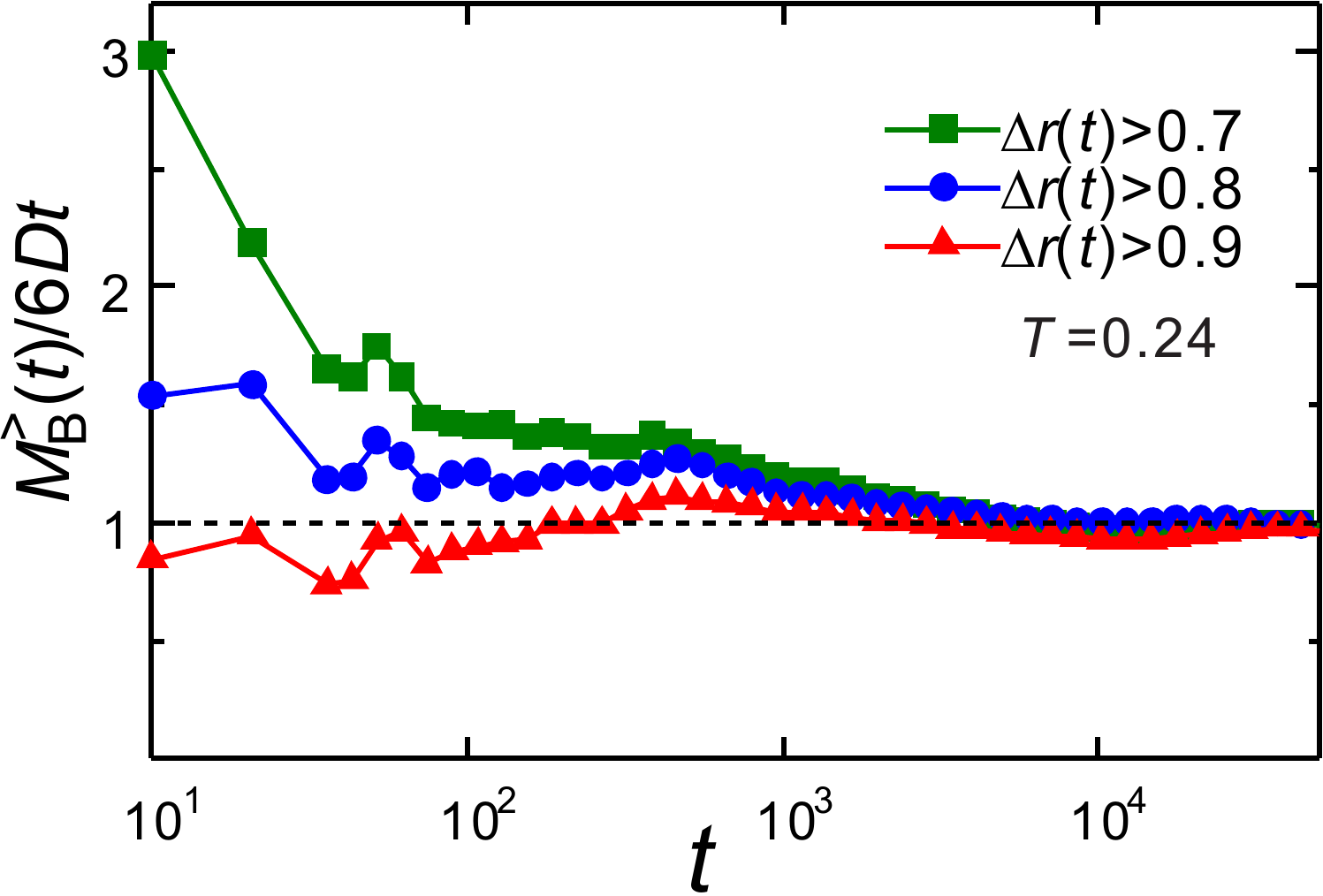}
\caption{(Color online)  $M_B^>(t)/6Dt$ from the $\bi{BL}$ particles with 
$r>\ell_{\rm m}=0,7, 0.8$, and $0.9$ for $T=0.24$, 
where $\ell_{\rm m}=0.8$ in the other figures in this paper. 
They  are close to the minimum distance  
$r_{\rm m}\sim 0.7$ of  $4\pi r^4 G(r,s)$ in Fig.9a. 
These three curves converge to unity for $t\gs 1000$. 
}
\end{figure}

\section{Summary and remarks}
\setcounter{equation}{0}

In this paper, simulations have been performed on a high-density 
binary mixture  at low $T$, where the particles   interact 
via  the soft-core potential (2.1) 
with the size ratio   $\sigma_2/\sigma_1=1.2$. 
The  total particle number  $N$  is  $10^4$ and the density $n$ is 
$0.8\sigma_1^{-3}$.  
Our main results are summarized  as follows.\\
(i)  In Sec.III,  using 
the   broken bond number 
 ${\cal  B}_i(t_0,t_1) $  in Eq.(3.5), 
 we have visualized the  particles 
composing strings and those surrounding them  in Fig.1. 
The former have undergone large displacements ($\Delta r_i>0.8)$ 
with multiple broken bonds (${\cal B}_i >1$), while the latter 
 small  displacements ($\Delta r_i<0.8)$ 
with a single  broken bond (${\cal B}_i=1$).
The number of the latter is several 
times larger than that of the former because of the large coordination 
numbers  in 3D, which leads to 
the difference of the fraction of 
the former $\phi_b^>(t)$ and that of the latter  $\phi_b^<(t)=\phi_b(t)- 
\phi_b^>(t)$ as in Fig.3. 
These two kinds of motions become increasingly distinct as $T$ is 
lowered  as in Figs.1, 2, and 4 and in Table 1.   
The latter contribution to  $F_s(q,t)$ 
in Eq.(2.4) becomes long-lived 
at low $T$ as illustrated in Fig.5. This  leads to the  steeper rise of 
$\tau_\alpha$ than the other characteristic times related to the bond breakage 
at low $T$ in Fig.5d.\\
(ii) 
In Sec.IV, we have found that  the contribution  to 
the mean square displacement $M(t)$ from the particles 
with large displacements  behaves as $6Dt$ soon after the ballistic 
regime.   The origin of 
the violation of the Stokes-Einstein relation 
has been ascribed to the steep growth of $\tau_\alpha (\propto \eta)$ 
at low $T$ for our system.   
\\

We further give   some remarks in the following.\\ 
(1)  
In our fragile glass-former,  the non-Arrhenius 
behavior in the Angell plot 
(the steep growth in  the curve of $\log \eta$ vs $1/T$) \cite{Angel}  
and the violation of the Stokes-Einstein relation \cite{Silescu,Ediger}
are  closely related under the condition $\tau_\alpha \sim \eta/T$. 
Note that we have not calculated  $\eta$ but 
assumed $\eta\propto T\tau_\alpha$ from  Eq.(2.6). 
In future,  we should calculate $\eta$  together with $D$  
 for  lower $T$.\\  
(2) In a 50:50 binary mixture with  the soft-core potential, 
we have examined  the particle motions 
around   strings.  We should  examine whether our results   
remain  valid or need to be modified  for other compositions and 
for other particle interactions \cite{Rei,Starr}. We should note that 
a 20:80 mixture with a 
Lennard-Jones potential was used to detect strings 
in the original work\cite{Kob}.  Stringlike collective motions 
were also observed in polycrystalline systems 
numerically \cite{Hamanaka,Jack} and experimentally \cite{Mary}.\\ 
(3)   
The   diffusion behavior 
of  the mean square displacement $M(t)$ 
can  be obtained  soon after the ballistic regime  
if  the contribution from large displacements 
are picked up. This is a natural result in 
 the activated dynamics.  For  lower $T$, 
   the   diffusion constant $D$  is   obtainable 
in this manner, where   its   aging behavior will  be of interest.

\begin{acknowledgments}
This work was supported by Grant-in-Aid 
for Scientific Research  from the Ministry of Education, 
Culture,  Sports, Science and Technology of Japan.  
 T. K. was supported by the Japan Society for Promotion of Science.
The authors would like to thank   
Hayato Shiba,  Kunimasa Miyazaki, and Kang Kim  
for informative discussions. 
The numerical 
calculations were carried out on SR16000 at YITP in Kyoto
University. 
\end{acknowledgments}

\end{document}